\documentclass[amsmath,amssymb,prc,final,showpacs,twocolumn]{revtex4} 

\usepackage{graphicx,epsfig}
\usepackage{times,epsfig,color,rotating,fancybox,pstricks,pst-coil,
                        bm,amsmath,amssymb,pst-plot}

\newcommand{\cm}{\mathrm{c\!\:\!.m\!\:\!.}}

\begin{document}

\title { Three-nucleon system:  Irreducible and reducible  contributions of
the three-nucleon force}

\author{A.~Deltuva} 
\affiliation{Institute of Theoretical Physics and Astronomy,  Vilnius
  University, A. Go\v{s}tauto 12, LT-01108 Vilnius, Lithuania}
\author{P. U. Sauer} \affiliation{Institute for Theoretical Physics,
  Leibniz University Hannover, D-30167 Hannover, Germany}

\received{\today}


\begin{abstract}
The three-nucleon bound and scattering equations are solved in
momentum space for a coupled-channel Hamiltonian. The Hamiltonian
couples the purely nucleonic sector of Hilbert space with a sector in
which one nucleon is excited to a $\Delta$ isobar. The interaction
consists of  irreducible two-baryon and irreducible three-baryon
potentials. The calculation keeps only the purely nucleonic one among
the irreducible three-baryon potentials. The coupled-channel
two-baryon potential yields additional reducible contributions to the
three-nucleon force. The Coulomb interaction between the two protons
is included using the method of screening and
renormalization. Three-nucleon force effects on the bound-state
energies and on observables of elastic nucleon-deuteron scattering and
breakup are studied.
\end{abstract}

 \maketitle

\section{ Motivation \label{sec:intro}}

The notion of the three-nucleon (3N) force is not a uniquely-defined
concept for the theoretical description of nuclear phenomena. The 3N
force arises in the hadronic picture of nuclear systems, a model
developed by theoreticians for calculational convenience. The 3N force
is therefore also model-dependent and experimentally not measurable. 

The microscopic degrees of freedom underlying nuclear phenomena are
the quarks and gluons of quantum chromodynamics (QCD). However, a
direct description of nuclear phenomena in terms of those microscopic
degrees of freedom is not available yet; instead, the effective
description in terms of quark-gluon clusters, i.e., in terms of
hadrons and their interactions among each other and with electroweak
probes is a common  conceptual and often quantitatively successful
 approach to nuclear phenomena 
at low and moderate energies  which we also adopt. At low energies, 
rigid nucleons appear to make up nuclei
in bound and scattering states; after all, nuclear bound states have
masses which are almost multiples of the one of a single nucleon. At
energies above the pion ($\pi$)-production threshold, the $\Delta$
isobar and the $\pi$ meson become additional active degrees of
freedom. The interactions between the hadronic constituents of nuclear
systems depend on the chosen degrees of freedom, i.e., on the energy
range of applicability; they are mediated by exchanged mesons. The
dynamics is usually   assumed to be controlled by a field theory for
the effective hadronic degrees of freedom, originally by a
phenomenological field theory with a zoo of mesons
\cite{machleidt:87a}; at present, the favorably employed one is the
chiral effective field theory ($\chi$EFT)
\cite{epelbaum:06,epelbaum:09,machleidt:11a}, which respects the chiral
symmetries of QCD and works with the nucleon and $\pi$ as degrees of
freedom, sometimes extended by the inclusion of the $\Delta$ isobar
\cite{krebs:07,krebs:08}. 

However for practical calculations of
nuclear systems, the dynamics is chosen to follow quantum mechanics.
The forces between the hadrons, arising from field-theoretic
processes, have therefore to be cast into the form of hermitian
potentials. That dynamic simplification is achieved by freezing some
of the field-theoretic degrees of freedom. The quantum-mechanical
kinematics is non-relativistic  for the nucleon and the $\Delta$
isobar, relativistic for the $\pi$, if quantum-mechanically
active. Quantum mechanics is advantageous, since it allows the step
from two-particle to many-particle systems in a natural way. The
potentials of the baryons are of two-, three- and, in general,
of A-particle nature, A being the number of baryons making up the
nuclear system under study. It is well established that at normal
densities the many-N potentials show a hierarchy of importance: The 2N
part is most important and makes the contact with the deuteron and
with free two-nucleon scattering. The 3N part is the first important dynamic
correction of the 2N interaction, whereas many-N interactions are more
complex than the 3N one and also appear dynamically rather unimportant
at normal nuclear densities. Furthermore, many-baryon contributions to
the interaction, e.g., those arising from non-nucleonic baryons as the
$\Delta$ isobar,  are also dynamically strongly suppressed due to their
reduced weights in the nuclear wave functions. That is the reason why
the present paper focuses on the 3N force in the hadronic picture of
nuclear phenomena. 

\begin{figure}[!]
\centerline{\includegraphics[width=0.4\textwidth]{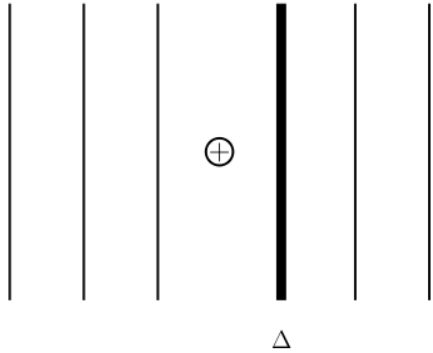}}
\caption{\label{fig:ExtendedHilbertSpace} Hilbert space for the
  description of nuclear phenomena at energies below the
  $\pi$-production threshold. The shown example is for baryon number
  3. Compared with a purely nucleonic Hilbert space, it is extended by
  a sector, in which one nucleon (thin line) is turned into a $\Delta$
  isobar (thick line). The deuteron and two-nucleon reactions  are
  described in the corresponding Hilbert space of baryon number 2.}
\end{figure}

The present paper makes the following choice for the description of
nuclear phenomena. The only active degrees of freedom are the nucleon
and the $\Delta$ isobar, all mesonic degrees of freedom are
frozen. The Hilbert space consists of two sectors, a purely nucleonic
one and a sector in which one nucleon is replaced by a $\Delta$
isobar. The Hilbert space is chosen with a view on the energy domain
above the $\pi$-production threshold. The $\Delta$ isobar is an
important mechanism for  $\pi$ production, but since 2$\pi$ and 3$\pi$
production are strongly suppressed far above their thresholds,
configurations with more than one $\Delta$ isobar should be physically
less important in the energy regime up to  about $ \rm 0.5~GeV$ excitation
in the c.m. system. Our goal is the description of nuclear phenomena
for energies below the  $\pi$-production threshold. Though the $\pi$
is energetically not an active degree of freedom yet, the $\Delta$
isobar is already expected to make important contributions to the
nuclear dynamics. That will indeed be the case, and that is the reason
for the chosen extended Hilbert space. The chosen Hilbert space is
shown in Fig.~\ref{fig:ExtendedHilbertSpace} for the example of a
three-baryon system.  In the presently chosen absence of a coupling to
$\pi$-nucleon states the $\Delta$ isobar is a stable baryon; its rest
mass is taken to be 1.232 GeV, the resonance position of  $\pi$-nucleon
scattering in the $\rm P_{33}$ partial wave.

\begin{figure}[!]
\centerline{\includegraphics[width=0.4\textwidth]{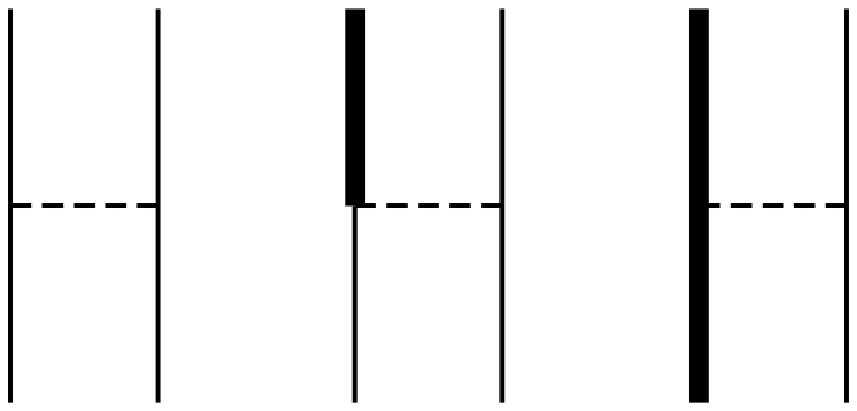}}
\centerline{\includegraphics[width=0.49\textwidth]{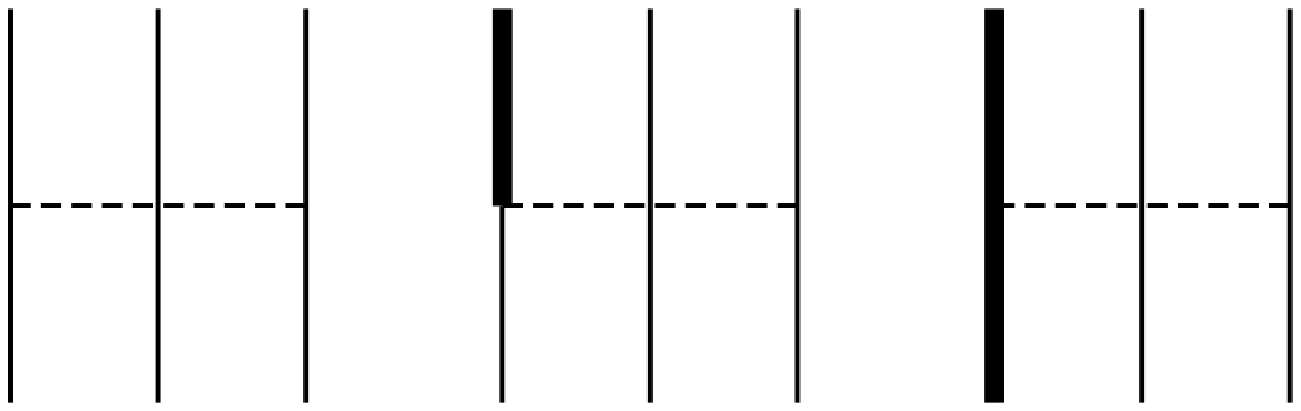}}
\caption{\label{fig:Hamilton} Interactions describing the nuclear
  dynamics in the Hilbert space of
  Fig.~\ref{fig:ExtendedHilbertSpace}. They are irreducible
  coupled-channel potentials of two-baryon (upper row) and
  three-baryon (lower row) nature. Besides their purely nucleonic
  parts, they couple the purely nucleonic sector with the one
  containing a $\Delta$ isobar, and they act directly in the sector
  containing a $\Delta$ isobar. Only selected examples of the
  potentials are shown; the hermitian-conjugate pieces have to be
  added as well as those obtained by permuting the $\Delta$ isobar in
  the initial and final states to other positions in the diagrams.}
\end{figure}

The potentials, required for the dynamics in the chosen Hilbert space,
are illustrated in Fig.~\ref{fig:Hamilton}; they are irreducible and
hermitian. The interaction Hamiltonian is of coupled-channel
character: It consists of two-baryon parts which act in both sectors
of the Hilbert space and which couple them by transition
potentials. The Hamiltonian also contains three-baryon parts which
also act in both sectors of the Hilbert space and couple them by
transition potentials. In addition there are  more complex
many-baryon potentials involving more than three baryons, up to baryon
number A, the baryon number of the nuclear system under
study. However, we shall constrain our calculations to the
three-nucleon system; thus, four-baryon forces and forces of even higher
complexity do not occur.

\begin{figure}[!]
\centerline{\includegraphics[width=0.4\textwidth]{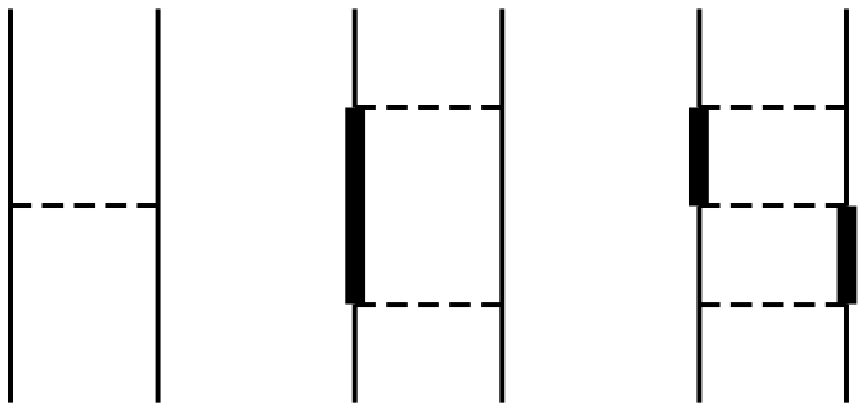}}
\centerline{\includegraphics[width=0.49\textwidth]{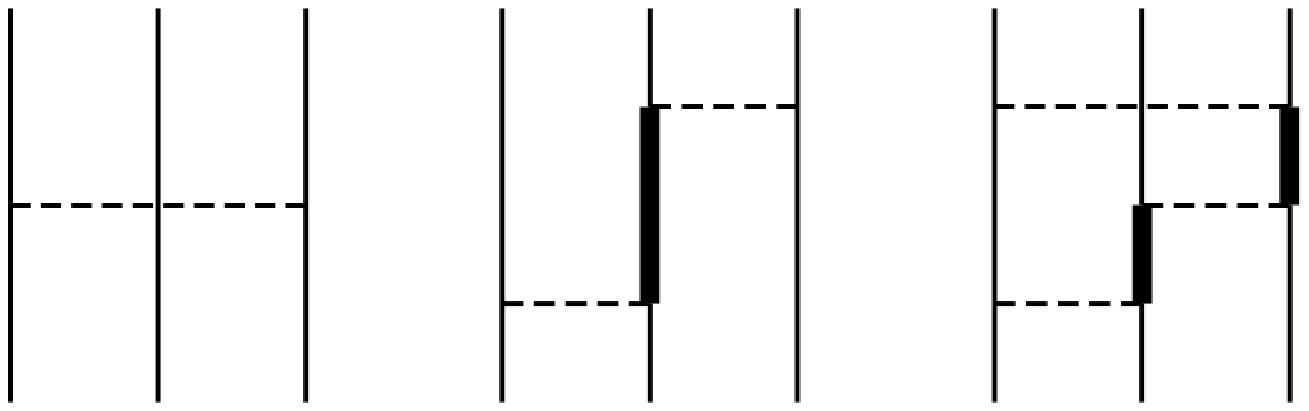}}
\caption{\label{fig:IrreducibleReducible} Irreducible contributions
  and selected examples of reducible contributions to the 2N (upper
  row) and to the 3N (lower row) interactions. The two reducible 3N
  processes  shown are, in sequence, the Fujita-Miyazawa and the
  $\Delta$-ring processes.}
\end{figure}

When employing the coupled-channel Hamiltonian with an explicit
$\Delta$ isobar, illustrated in Fig.~\ref{fig:Hamilton}, the nuclear
dynamics receives the irreducible contributions of the chosen
Hamiltonian, but also reducible contributions to the 2N and 3N
interactions arising from its iterative application in the process of
solving the many-nucleon problem. The interplay between irreducible and
reducible contributions to the 2N and 3N interactions is shown in
Fig.~\ref{fig:IrreducibleReducible}. Some dynamic processes, in other
approaches  described by irreducible potentials, are already accounted
for by the coupled-channel two-baryon part of the  Hamiltonian as reducible 
processes in the Hilbert space of
Fig.~\ref{fig:ExtendedHilbertSpace}; they have therefore to be
excluded from the irreducible parts of the
Hamiltonian. 

The 2N interaction receives an important contribution from the virtual
excitation of a nucleon to the $\Delta$ isobar contributing to the
attraction at intermediate range. For example, the potential CD Bonn + $\Delta$,
designed by us as such a coupled-channel potential in Ref.~\cite{deltuva:03c}
and employed by us for the description of few-nucleon systems previously,
artfully takes out the virtual excitation of the $\Delta$ isobar from
its irreducible part. The reducible, usually attractive contributions,
illustrated in the upper row of  Fig.~\ref{fig:IrreducibleReducible},
to the 2N interaction get weakened in the nuclear medium,  a
well-known effect, called 2N dispersion.

In the same way, the three-nucleon processes provided by the coupled-channel
potential in a reduced form will have to be excluded  from the
employed irreducible 3N potential. The physically most important
three-nucleon processes which the coupled-channel potential already
provides are the Fujita-Miyazawa~\cite{fujita:57a} and the $\Delta$-ring
processes, as illustrated in the lower row of
Fig.~\ref{fig:IrreducibleReducible}; the importance of those processes
is seen in our own calculations, but it is also confirmed in the context of other
dynamic strategies in Refs.~\cite{krebs:13,pieper:01a}. 

The paper studies the three-nucleon system. Its objective is twofold,
it has a physics aspect and a technical aspect:
\begin{itemize}
\item
The paper extends the dynamics of our previous calculations~\cite{nemoto:98a,deltuva:03a}
by adding an irreducible 3N force to the employed coupled-channel two-baryon
potential. Our previous calculations without such an irreducible 3N
force are physically incomplete and missed the three-nucleon binding energies and
the neutron-deuteron doublet scattering length. We would like to
improve on both quantities with the physics question: Is their simultaneous
improvement possible? How serious is the miss on those two observables
for the prediction of three-nucleon scattering at higher energies?
\item
The paper demonstrates that complete and reliable calculations
including reducible and irreducible 3N forces can be done for the
bound and scattering states of the three-nucleon system. As will be
discussed in Subsection II.B, the chosen dynamics is not yet in all its
respects up to the most modern standards. But the paper paves the way
to further,  dynamically more satisfactory calculations. That
is the technical objective of this paper.
\end{itemize}


\section{ Calculational apparatus} 

The paper calculates the three-nucleon bound and scattering states,
i.e., nucleon-deuteron elastic scattering and breakup, for energies up to
the $\pi$-production threshold. Those calculational goals require the
extension of two  numerical schemes, used previously by the
authors. Our standard approach to the three-nucleon system is based on
a coupled-channel  two-baryon potential as in 
Refs.~\cite{nemoto:98a,deltuva:03a};  note that Ref.~\cite{deltuva:03a} 
and all our recent calculations overcame the need for a separable 
expansion of the two-baryon transition matrix, still required in 
Ref.~\cite{nemoto:98a}.
In contrast, Ref.~\cite{deltuva:09e} uses irreducible
two- and three-baryon forces for the three-nucleon bound states and scattering,
but the forces are purely nucleonic ones.  The present paper requires
the combination of those two schemes.  Ref.~\cite{stadler:92a} already
 implemented such a combination of a coupled-channel two-baryon potential
with an explicit $\Delta$-isobar and of an additional irreducible 3N force, but the
application was constrained  to the three-nucleon bound state; the
present paper extends that approach to the three-nucleon
scattering. Furthermore, the coupled-channel potential, employed in
Ref.~\cite{stadler:92a}, was not properly fitted to the two-nucleon
data. 

\subsection{ The equations}

There are different strategies for setting up the quantum-mechanical
equations for obtaining the three-nucleon bound and  scattering
states.  We work in momentum-space representation and  employ
integral equations  of the Alt, Grassberger, and Sandhas (AGS) type
\cite{alt:67a}. There are three pairwise potentials  $v_\alpha$, i.e., 

\begin{equation} 
 V_{(2)} = \sum_{\alpha=1}^3 v_\alpha,
\end{equation}

\noindent $\alpha$ being the label of the non-interacting particle
during the pairwise interaction. Each of the three potentials is
summed up into the respective two-particle transition matrix
$T_\alpha(z)$, i.e., 

\begin{equation}  
 T_\alpha(z) =  v_\alpha + v_\alpha G_0(z) T_\alpha(z).
\end{equation}

\noindent $G_0(z) = (z - H_0)^{-1}$ is the free three-particle
resolvent with the available energy $z$ and the free Hamiltonian $H_0$
containing the kinetic energy of the particles and their rest mass
differences, in the three-nucleon system with $\Delta$-isobar states
the difference with respect to three nucleons.  
The three-particle
potential 
\begin{equation} \label{eq:V3}
 V_{(3)} = \sum_{\gamma=1}^3 u_\gamma
\end{equation}
is also decomposed into three terms $u_\gamma$ that 
are symmetric under the exchange of particles 
$\beta \ne \gamma$ and $\alpha \ne \gamma$ 
and can be transformed into one another by cyclic permutations.
{
The decomposition (\ref{eq:V3}) is not unique. Often the 3N force is written
such that  $\gamma$ labels the particle which is directly
linked to the two others in the interaction \cite{urbana9}.
However, such a decomposition is inefficient for our practical calculation,
and we will therefore follow the procedure of Refs.~\cite{lazauskas:phd,deltuva:09e}; 
this procedure for the decomposition gets explained in Subsection II.B.
}

The strategy for setting up the integral equations works with the
two-particle part of the potential summed into the transition matrix
$T_\alpha(z)$, but keeping the three-particle part in its potential
form $V_{(3)}$ uniterated. The basis of states,  employed for
calculations in the three-nucleon system, is, in the  purely nucleonic
Hilbert sector, the direct product of an antisymmetrized  nucleon pair
and a third nucleon without antisymmetrization; thus, 
the basis is overcomplete.
  In the Hilbert sector
with a $\Delta$ isobar the same pair symmetrization  is used, since
the $\Delta$ isobar can be created from any of the three  nucleons in
a symmetric fashion. The isospin formalism is used for  distinguishing
protons and neutrons including both total isospin $\frac12$ and $\frac32$ 
states.   The results of the
integral equations are bound-state amplitudes $ |\psi_\alpha\rangle$
and multichannel transition amplitudes $U_{\beta \alpha}$, connecting
the initial two-body channel with the non-interacting particle of label
$\alpha$ to the final two-body channel with the non-interacting
particle of label $\beta$, $\beta = 0$ being the final three-particle breakup
channel. All multichannel transition amplitudes depend on the
available energy $E + i0$; that dependence is notationally
suppressed. The derivation of the equations is given in
Ref.~\cite{deltuva:09e}; the fact, that Ref.~\cite{deltuva:09e}
describes the three-nucleon system in terms of a completely nucleonic
Hilbert space is immaterial for the derivation of equations; the
resulting equations of Ref.~\cite{deltuva:09e} can therefore be taken
over without any change.

\subsubsection{ Bound state }

The three-particle bound-state wave function  
\begin{equation}
|\Psi \rangle = \sum_\alpha |\psi_\alpha  \rangle
\end{equation}
is decomposed into its Faddeev components $ |\psi_\alpha\rangle$.
Given the identity of nucleons, carried over to states with a $\Delta$
isobar,  the three Faddeev components are related by the permutation
of the baryons. Any Faddeev component can be taken as a symmetrized
Faddeev amplitude $|\psi \rangle$ which is obtained from
\begin{gather} 
|\psi  \rangle = G_0 T P |\psi \rangle + (1+G_0 T) G_0  u (1+P) |\psi
\rangle,
\end{gather}
yielding the bound-state wave function in the form  $|\Psi \rangle
=(1+P)|\psi \rangle$  with the help of the permutation operator $P$, i.e., $P =
P_{12}P_{23} + P_{13}P_{23}$, $P_{\beta \alpha}$ being  the individual
permutation operator of baryons $\beta$ and $\alpha$. 

\subsubsection{ Scattering operators }

The  multichannel transition amplitudes $U_{\beta \alpha}$ act on the
two-body channel states $|\phi_{\alpha}\rangle$, the eigenstates of
the corresponding channel Hamiltonian $H_\alpha = H_0 + v_\alpha$ with
the energy eigenvalue $E$. Due to symmetry, all three channel states
are related to each other by the permutation $P$, defined in the
previous subsection. One channel can therefore be taken as
characteristic for all. Its channel label will therefore be dropped as well as the label for the component $u_\gamma$ of the three-nucleon potential,
and one can work with a symmetrized transition operator $U$ for
elastic scattering, obtained from the symmetrized AGS integral
equation, i.e.,

\begin{subequations} 
\begin{gather} 
\begin{split}
U = {} & PG_0^{-1} + (1+P)u + PTG_0U \\ & + (1+P)uG_0(1+TG_0)U,
\end{split}
\end{gather}

\noindent yielding the on-shell matrix elements $\langle \phi|U|\phi
\rangle$, required for the calculation of observables.  The breakup
operator $U_0$, connecting an initial two-body channel with the final
breakup channel $|\phi_0 \rangle$ of non-interacting nucleons is then
obtained by quadrature

\begin{gather} 
U_0 = (1+P)[G_0^{-1} +u +TG_0U + uG_0(1+TG_0)U],
\end{gather}
\end{subequations} 

\noindent yielding the on-shell matrix elements $\langle
\phi_0|U_0|\phi \rangle$, required for the calculation of breakup observables.

In the following 
elastic scattering observables are shown as functions of the center-of-mass (c.m.) scattering angle $\Theta_\cm$,
while breakup observables  as functions of the arclength $S$ along the 
kinematical curve. 
As in Ref.~\cite{deltuva:09e},
we obtain well-converged results by taking into account
the hadronic interaction in two-baryon partial waves with the
total pair angular momentum $j_x \le 5$ and including three-baryon states
with total $3N$ angular momentum  $\mathcal{J} \le \frac{59}{2}$; 
it is fully sufficient to limit  the irreducible $3N$ force to states with
$\mathcal{J} \le \frac{19}{2}$. In all calculations of Subsection III.B for 
the three-nucleon system with two protons their
Coulomb repulsion is added with the necessary changes in the above
equations according to Refs.~\cite{taylor:74a,alt:80a,deltuva:05a,deltuva:05d}.

\subsection{Choice of the dynamics} 

We take over the potential CD Bonn + $\Delta$ \cite{deltuva:03c} as
coupled-channel two-baryon potential, keeping its parameters
unchanged in this paper. It is based on the exchange of $\pi$, 
$\rho$, $\omega$, and $\sigma$ mesons. 
We remind that this potential is tuned to the deuteron and
to two-nucleon elastic scattering below 350 MeV lab energy,
and that the potential parts, connected with the $\Delta$ isobar, are
therefore underdetermined by that procedure. Those potential parts
would be physically better determined by a simultaneous fit to
inelastic two-nucleon scattering, i.e., to the data of channels with a
single $\pi$. However, at present such a tuning procedure is too
demanding for us; it would also require the addition of a Hilbert
sector with a single $\pi$ and  the corresponding additions to the
Hamiltonian part of baryon number 2.  

We add an irreducible 3N potential to the Hamiltonian. Among the
multitude of possible three-baryon potentials, illustrated in the lower row of
Fig.~\ref{fig:Hamilton}, the 3N one was argued by us in Section I to be most
important.

We choose the Urbana IX 3N force \cite{urbana9} as starting point
for our  construction of 3N potential models, employed in this
paper. This choice of a phenomenological irreducible 3N potential is
against our original philosophy when developing a coupled-channel
two-baryon potential: 2N and 3N forces have to be physically
consistent with each other. The coupled-channel approach creates such
a satisfying consistency between the 2N and the reducible 3N
interactions; but this consistency is broken by the addition of a
phenomenological irreducible 3N potential, as done now. The Urbana IX
3N force, chosen as starting point for the modeling in the present
paper, has $2\pi$-exchange and phenomenological repulsive short-range
terms, i.e.,
\begin{gather} \label{eq:U9}
\begin{split}
V_{(3)} =  {} & \sum_{\alpha\beta\gamma \; \mathrm{cyclic}} \big(
A_{2\pi} \{X^\pi_{\alpha\beta},X^\pi_{\beta\gamma}\} \{\tau_\alpha
\cdot \tau_\beta,\tau_\beta  \cdot \tau_\gamma \} \\ & +  C_{2\pi}
      [X^\pi_{\alpha\beta},X^\pi_{\beta\gamma}] [\tau_\alpha \cdot
        \tau_\beta ,\tau_\beta  \cdot \tau_\gamma ]  + B_0
      T^2_{\alpha\beta} T^2_{\beta\gamma} \big),
\end{split}
\end{gather}
\noindent where curly and square brackets denote anticommutators and
commutators, respectively. The strength constants were chosen in
Ref.~\cite{urbana9}  to be $A_{2\pi} = -0.0293$ MeV, $C_{2\pi} = \frac14 A_{2\pi}$, 
and $B_0 = 0.0048$ MeV. 
Here $\tau_\beta$ is the
isospin operator of the nucleon $\beta$, while the potential-like pair $(\alpha\beta)$
operators $X^\pi_{\alpha \beta}$ and  $T_{\alpha \beta}$ are given in coordinate space by
\begin{subequations} \label{eq:U9x}
\begin{align} 
X^\pi_{\alpha\beta} = {} & Y(m_\pi r_{\alpha\beta}) \sigma_\alpha \cdot \sigma_\beta 
+ T(m_\pi r_{\alpha\beta}) S_{\alpha\beta} , \\
T_{\alpha\beta} =  {} & T(m_\pi r_{\alpha\beta}),
\end{align}
\end{subequations}
with $\sigma_\alpha$ being the spin operator of the nucleon $\alpha$,
$S_{\alpha\beta}$ the tensor operator of the pair ${\alpha\beta}$,
$m_\pi$ the average $\pi$ mass, and $Y(x)$ and $T(x)$ the Yukawa and
tensor functions specified in Ref.~\cite{wiringa:95a}; 
all operators are of $\pi$ range. 
In our calculation $X^\pi_{\alpha\beta}$ and  the squared operators $T^2_{\alpha\beta}$ are transformed to 
momentum space using spherical Bessel functions. 
 Since the decomposition of $V_{(3)}$ into three components according
 to Eq.~(\ref{eq:V3}) is not unique, we make the following choice:
We  expand  the sum (\ref{eq:U9}) 
and take as $u_\gamma$ the terms where the operator acting on the
final state (the operator that is on the left side in the product)
is either $X^\pi_{\alpha\beta}$ or $T^2_{\alpha\beta}$, i.e.,
\begin{gather} \label{eq:u9}
\begin{split}
u_\gamma =  {} &  2 X^\pi_{\alpha\beta} \big\{
X^\pi_{\beta\gamma} [  A_{2\pi} \tau_\alpha \cdot \tau_\gamma - 
i C_{2\pi} \tau_\alpha \cdot (\tau_\beta \times \tau_\gamma) ] \\ & +
X^\pi_{\gamma\alpha} [A_{2\pi}  \tau_\gamma \cdot \tau_\beta  + 
i C_{2\pi}  \tau_\alpha \cdot (\tau_\beta \times \tau_\gamma)] \big\} \\ &  +
\frac12 { B_0} T^2_{\alpha\beta} (T^2_{\beta\gamma} + T^2_{\gamma\alpha}).
\end{split}
\end{gather}
$V_3$ gets then decomposed as in Eq.~(\ref{eq:V3}). This decomposition is more efficient
in our practical calculations as argued in Refs.~\cite{lazauskas:phd,deltuva:09e}.
The numerical technique is taken over from Ref.~\cite{deltuva:09e} and adapted
here for the context of additional non-nucleonic channels. This
technical possibility of casting the 3N potential $V_3$ into the form ~(\ref{eq:V3}) with $u_\gamma$ of Eq. (\ref{eq:u9}) was partly the reason for choosing the Urbana IX 3N
force as basis  for our search of a useful irreducible 3N force
supplementing a given coupled-channel two-baryon potential. 

The $A_{2\pi}$ and $C_{2\pi}$ terms in $u_\gamma$ and $V_3$ are related to $2\pi$-meson
exchange; the Fujita-Miyazawa mechanism, illustrated in its
reducible form in Fig.~\ref{fig:IrreducibleReducible}, makes an
important contribution to the 3N potential, requiring the parameter
combination $C_{2\pi} = \frac14 A_{2\pi}$, 
as chosen in Ref.~\cite{urbana9}. Since that dynamic part is
provided by the employed coupled-channel two-baryon potential CD
Bonn+$\Delta$ in a reducible form, the Fujita-Miyazawa contribution
has to be taken out from the parameters in our reparametrization of an
irreducible 3N potential in the Urbana-like fashion. We note in passing, 
that the reducible Fujita-Miyazawa process, provided by the coupled-channel two-baryon 
potential CD Bonn + $\Delta$, also includes the $\rho$-meson exchange and is 
therefore richer than the original Fujita-Miyazawa 3N force of Ref.~\cite{fujita:57a}. 
Compared to the
Urbana IX 3N force the strength of the potential should be reduced and
deviations from the ratio $C_{2\pi}/A_{2\pi} = \frac14 $ are
acceptable. In fact, that  ratio $C_{2\pi}/A_{2\pi}$  
got increased in Ref.~\cite{stadler:92a} when the Fujita-Miyazawa 
process was removed there from the Tucson-Melbourne 
3N potential~\cite{coon:79}. We therefore have to expect a similar 
increase of that  ratio $C_{2\pi}/A_{2\pi}$ 
in our reparametrization of an irreducible 3N potential in the Urbana-like fashion.

{ The repulsion with the strength parameter $B_0$ in the
Urbana IX 3N force was required to prevent overbinding of nuclear matter. 
However, that argument is not relevant in the present context any longer. 
The coupled-channel two-baryon potential provides repulsion due to the two-nucleon 
dispersion. That fact was confirmed for nuclear matter by Ref.~\cite{manzke:78}.
We therefore expect that our
reparametrization of the irreducible 3N potential should have
a smaller or even a vanishing $B_0$ value compared to the
Urbana IX 3N force. We therefore set $B_0 = 0$ in our first try
and retune $A_{2\pi}$ and $C_{2\pi}$  having in mind }
that the resulting dynamic model has to account for the $^{3} \rm H$
binding energy $ |E_{^{3} \rm H}|$
and attempting also to fit the  neutron-deuteron doublet
scattering length $a_2$. The potential CD Bonn + $\Delta$ misses $ |E_{^{3} \rm H}|$
only by 176 keV and with $a_2 = 0.695$ fm is  already pretty close to the
experimental value of 0.65$\pm $0.04 fm \cite{dilg:71}. Thus, the required parameters in
the added modified Urbana-like 3N potential should be quite mild ones.

However, with the coupled-channel two-baryon potential 
CD Bonn + $\Delta$ and the modified 
Urbana IX 3N force as irreducible 3N force in addition, we are unable
to  fit  the $^{3} \rm H$ binding
energy $ |E_{^{3} \rm H}|$ and  the neutron-deuteron doublet scattering length $a_2$
simultaneously  with weak $A_{2\pi}$ and $C_{2\pi}$ parameters.
We therefore included also the $B_0$ term and
tried a wide  range of parameters, using a rather coarse raster,
but staying within  sensible limits. In the end we chose
two groups of modifications  only, whose parameters are listed in the
Table \ref{tab:u} and which appear to us characteristic for our parameter 
search with Urbana-like 3N potentials.
In the first group the ratio  $C_{2\pi}/A_{2\pi}$ 
was set larger than $\frac14$ by hand and not subjected to the fit.

The first group of parametrizations consists of the 3N potentials (U1,
U2, U3). The two terms augmented by the parameters $A_{2\pi}$ and
$C_{2\pi}$ yield both attraction; if repulsion is needed, it is
provided by the term with the parameter $B_0$. 
All models fit  $ |E_{^{3} \rm H}|$ perfectly and,
except for U3,  underpredict the neutron-deuteron doublet scattering length $a_2$.
Running down this list
of three modifications (U1, U2, U3) the individual strength parameters
for attraction and repulsion become increasingly larger; when
accounting for the neutron-deuteron doublet scattering length $a_2$ as
in U3, the three parameters are individually largest. Furthermore, we
are disturbed that the $A_{2\pi}$ and $C_{2\pi}$ parts of the 3N
potential U3 make a massive contribution of about 5 MeV to the triton binding 
$ |E_{^{3} \rm H}|$, which has to be balanced by a corresponding
comparable repulsion due to the $B_0$ term, just to squeeze out the
required  small additional binding of 176 keV. 
Thus,  the simultaneous  account of the $^{3}\rm H$ binding $ |E_{^{3} \rm H}|$ and of
the neutron-deuteron  doublet scattering length
$a_2$ appears  difficult. We are only able to achieve that
goal  with the irreducible 3N potential U3 of questionable parametrization, i.e., the
attraction and repulsion required for balancing each other, are 
unnaturally large.  However, this tuning difficulty appears not to be a 
feature characteristic for our special 3N force model.  
Indeed, in Ref.~\cite{kievsky:10a} another study was performed  using the purely nucleonic 
Argonne V18 potential \cite{wiringa:95a} in combination with various 3N forces; 
 reproducing experimental values of the neutron-deuteron  doublet scattering 
length $a_2$ and of 3N and 4N binding energies required a balance of 
attractive and repulsive 3N force contributions, individually even stronger than in the model
CD Bonn + $\Delta$ + U3 of this paper.
Furthermore, Ref.~\cite{epelbaum:14} used purely nucleonic 2N and 3N potentials consistently derived from
$\chi$EFT, thus, it uses rather different dynamics; nevertheless, the simultaneous account of the $^{3}\rm H$ 
binding $ |E_{^{3} \rm H}|$ and of the neutron-deuteron 
doublet scattering length $a_2$ was also either impossible at all or led
to unnaturally large coupling constants, quite similar to our finding.
In addition, we note that all realistic purely nucleonic potentials combined 
with standard phenomenological 3N forces fitted to $ |E_{^{3} \rm H}|$ are
underpredicting $a_2$ by a comparable amount \cite{witala:03a}.

In some parameter domains we noted a surprising sensitivity for the
nucleon analyzing power $A_y$ of elastic nucleon-deuteron
scattering. We therefore created a second group of parametrizations
consisting of the 3N potentials (U1, U4, U5). The repulsive term with
the parameter $B_0$ is left out for them; attraction and, if needed, repulsion
are both provided by the  $A_{2\pi}$ and $C_{2\pi}$ terms. In the 3N
potentials U4 and U5  the two parameters have opposite signs. We
remind that also the 3N potential U1 has $B_0 = 0$, whereas the two
other parameters yield attraction; a 3N potential with both
parameters $A_{2\pi}$ and $C_{2\pi}$ yielding repulsion is unable to account 
for  the $^{3}\rm H$ binding $ |E_{^{3} \rm H}|$.
None of the parametrizations of the second group (U1, U4,
U5) is able to account for the neutron-deuteron doublet scattering
length $a_2$, but U5 is roughly adjusted to  the neutron
analyzing power $A_y$ of elastic neutron-deuteron scattering at 10
MeV neutron energy. 

Additional modifications with further intermediate parameters appear
unnecessary to be accounted in the paper, since the two groups of models for an irreducible 3N
force suffice to describe the observed trends. 

\begin{table}[htbp]
  \centering
  \begin{tabular}{l*{5}{c}}
&  $A_{2\pi}$  &  $C_{2\pi}/A_{2\pi} $  &  $B_0$  & $ |E_{^{3} \rm H}|$ 
&  $a_2$     \\ \hline   
CD Bonn & & & & 8.004 & 0.932    \\   
CD  Bonn+$\Delta$ & & & & 8.306 & 0.695 \\ 
\hline  
CD  Bonn+$\Delta$+U1  & -0.00266 & +0.5000 & 0 & 8.482 & 0.559    \\  
CD Bonn+$\Delta$+U2 &  -0.01559 & +1.0000 & 0.00625 & 8.482 &   0.606 \\  
CD Bonn+$\Delta$+U3 & -0.02598 & +1.0000 & 0.01250 &    8.482 & 0.639 \\ 
\hline  
CD Bonn+$\Delta$+U1 & -0.00266 & +0.5000     & 0 & 8.482 &    0.559 \\  
CD Bonn+$\Delta$+U4 & +0.06500 &     -0.3995 & 0 & 8.482 &    0.576\\  
CD Bonn+$\Delta$+U5 & -0.06500 &    -0.3551 & 0 & 8.482 &  0.539 \\

  \end{tabular}
\caption{ \label{tab:u} Parametrization of the employed irreducible 3N
  potentials.  The parameters  $A_{2\pi}$, $C_{2\pi}$  and $B_0$ refer
  to the form of the 3N potential of Eq.~(\ref{eq:U9}); they are given
  in units of MeV. The $^{3} \rm H$ binding energy $|E_{^{3}\rm H}|$ is given
  in MeV, its experimental value
  being 8.482 MeV;  the  neutron-deuteron doublet scattering length
  $a_2$ is given in fm, its experimental value being 0.65$\pm $0.04 fm.
  In contrast to other calculations in this work, the results are obtained including
 3N states with the 2N total angular momentum { $ j_x \le 6$}.
}
\end{table}

\section {Results \label{sec:res}} 
{
\subsection{Observations made in the parameter search}
}   
In this subsection, we compare the predictions of interactions with the added
irreducible 3N potentials, listed in the Table \ref{tab:u}, for selected
three-nucleon scattering observables; we do so separately for the two
distinct groups (U1, U2, U3) and (U1, U4, U5). We choose selected observables
which best illustrate the effects of the added 3N potentials in either
group, desired and undesired, i.e., the neutron analyzing power $A_y$ of 
neutron-deuteron elastic scattering at 10 MeV neutron energy, the differential 
cross section $d\sigma/d\Omega$
and the deuteron analyzing power  $A_y(d)$ of proton-deuteron elastic scattering 
at 135 MeV proton energy, and the differential cross section
for neutron-deuteron breakup at 13~MeV neutron lab energy  in the space star
configuration.

\begin{figure}[!]
\begin{center}
\includegraphics[scale=0.64]{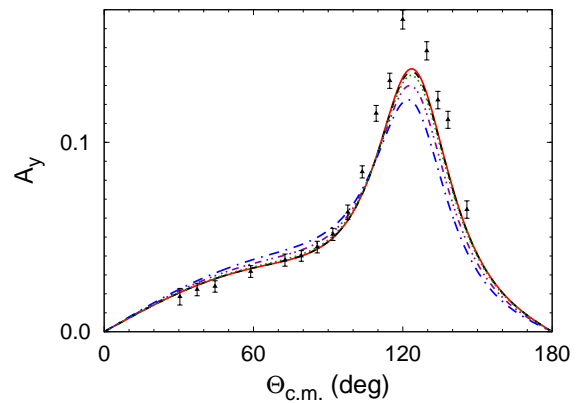}
\end{center}
\caption{\label{fig:nd10u3} (Color online) Neutron analyzing power
  $A_y$ for neutron-deuteron elastic scattering at  10~MeV neutron
  energy as function of the c.m. scattering angle $\Theta_\cm$.  Results of CD Bonn
  (dotted curve), CD Bonn + $\Delta$ (dashed curve),  CD Bonn +
  $\Delta$ + U1 (solid curve), CD Bonn + $\Delta$ + U2
  (double-dotted-dashed curve),  and  CD Bonn + $\Delta$ + U3
  (dotted-dashed curve) are compared with the  experimental data from
  Ref.~\cite{howell:87}.}
\end{figure}

\begin{figure}[!]
\begin{center}
\includegraphics[scale=0.64]{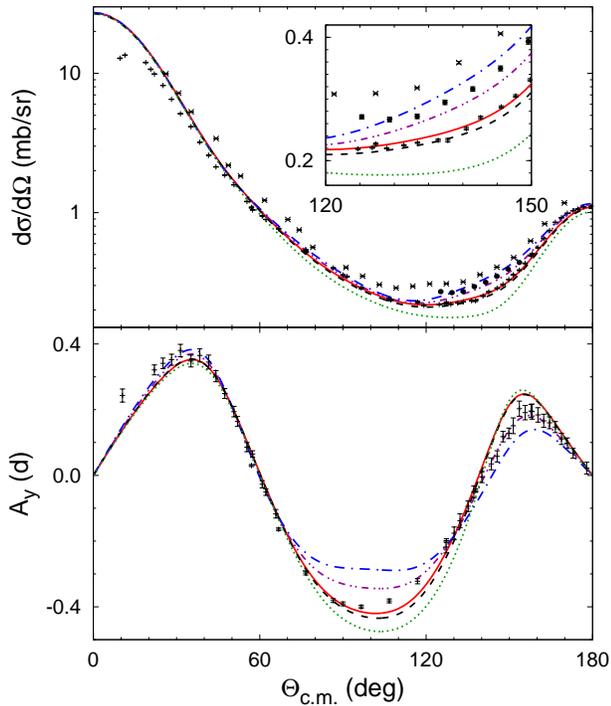}
\end{center}
\caption{\label{fig:nd135u3} (Color online) { Differential cross section
  and deuteron vector analyzing power for proton-deuteron elastic
  scattering at  135~MeV proton energy as function of the
  c.m. scattering angle $\Theta_\cm$.  Curves as in Fig.~\ref{fig:nd10u3}.  
The theoretical predictions of this figure are obtained omitting the Coulomb 
repulsion between the protons for reasons of calculational simplicity; 
at this energy Coulomb does not yield any 
noticeable effect at angles $\Theta_\cm > 20^\circ$~\cite{deltuva:05a}; 
its inclusion is therefore immaterial for the objectives of this subsection.  The
  experimental data are from Refs.~\cite{ermisch:05a} (x),
  \cite{ramazani:08a} ($\bullet$), and \cite{sekiguchi:02a} (+).}}
\end{figure}

{
First, results for group (U1, U2, U3) are illustrated in 
 Figs.~\ref{fig:nd10u3} and \ref{fig:nd135u3}. 
The general trend is that 3N force effects due to the irreducible U1 are 
insignificant, even smaller than those obtained from the reducible 3N force contributions 
due to the explicit $\Delta$-isobar excitation; they improve the agreement with the experimental data
\cite{howell:87,ermisch:05a,ramazani:08a,sekiguchi:02a} only slightly.
In contrast, the effects due to the 3N potential U3 are larger than those 
due to the explicit $\Delta$-isobar excitation, but not beneficial for the description 
of the data. At 135 MeV proton-deuteron elastic scattering, they move the theoretical predictions into the right direction, but they
are far too strong, often dramatically overshooting the discrepancy with data. This somehow uncontrolled behavior is not unexpected, given the individually very strong
repulsive and attractive terms in U3. The results for U2 are
intermediate between those for U1 and U3. We conclude that the
irreducible 3N potential U1, being the softest one,  appears to make
most sense physically, even though it predicts $a_2 = 0.559$ fm and
we have to give up our ambition to account for the experimental 
 $a_2$ value precisely, when choosing U1 for further studies. 

Concerning the potential CD Bonn + $\Delta$ + U3 that is forced to fit the
neutron-deuteron  doublet scattering length $a_2$, 
we observe another interesting 
similarity in Fig.~\ref{fig:nd10u3} with the  results of Ref.~\cite{epelbaum:14}, which are obtained
from $\chi$EFT 2N  and 3N forces, also fitted to the $^{3}\rm H$ binding $ |E_{^{3} \rm H}|$ and to the 
neutron-deuteron doublet scattering length $a_2$. In addition to the similarity found 
in Subsection II.B, we observe that in both cases, based on completely different dynamic models, the 
neutron analyzing power $A_y$ in neutron-deuteron elastic scattering at 10 MeV
is increased at angles $\Theta_\cm < 100^\circ$,
but decreased  $\Theta_\cm > 110^\circ$, moving theoretical predictions
away from the experimental data. Thus, this additional undesired effect of irreducible 3N forces
on the analyzing power $A_y$ appears to be a general feature of presently available force models
that  fit 2N data, $ |E_{^{3} \rm H}|$ and $a_2$ simultaneously and points to a further aspect of 
the long-standing $A_y$ puzzle in low-energy 3N scattering.
 In Ref.~\cite{kievsky:10a} the analyzing power was studied only at 3 MeV, but the conclusions
are essentially the same, i.e., improving $a_2$ increases the discrepancy for $A_y$, except 
when using the Argonne V18 potential supplemented with  a particular 3N 
force of $\chi$EFT.

\begin{figure}[!]
\begin{center}
\includegraphics[scale=0.64]{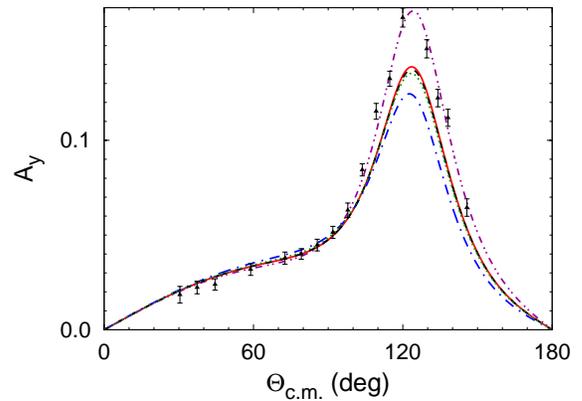}
\end{center}
\caption{\label{fig:nd10u5} (Color online) Neutron analyzing power
  $A_y$ for neutron-deuteron elastic scattering at  10~MeV neutron
  energy as function of the c.m. scattering angle $\Theta_\cm$.  Results of CD Bonn
  (dotted curve), CD Bonn + $\Delta$ (dashed curve),  CD Bonn +
  $\Delta$ + U1 (solid curve) CD Bonn + $\Delta$ + U4 (dotted-dashed
  curve) and  CD Bonn + $\Delta$ + U5 (double-dotted-dashed curve) are
  compared with the  experimental data from Ref.~\cite{howell:87}.}
\end{figure}

\begin{figure}[!]
\begin{center}
\includegraphics[scale=0.64]{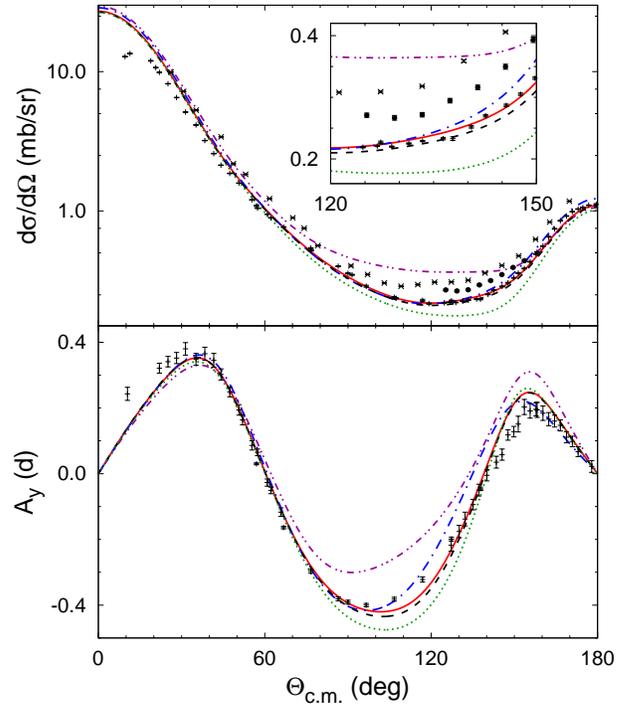}
\end{center}
\caption{\label{fig:nd135u5} (Color online) Differential cross section
  and deuteron vector analyzing power for proton-deuteron elastic
  scattering at  135~MeV proton energy as function of the
  c.m. scattering angle $\Theta_\cm$.  
 The theoretical predictions of this figure are obtained omitting 
the Coulomb force for reasons explained in the caption of 
Fig.~\ref{fig:nd135u3}. Curves as in Fig.~\ref{fig:nd10u5}.
 The experimental data are quoted in Fig.~\ref{fig:nd135u3}.}
\end{figure}

Second, the above conjecture on the general difficulty of accounting simultaneously 
for characteristic  three-nucleon data, seen for the potential group (U1, U2, U3), 
is supported also by our 
results for the potential group (U1, U4, U5) as shown in Fig.~\ref{fig:nd10u5}.
The model  CD Bonn + $\Delta$ + U5 is adjusted to reproduce the experimental data
for the neutron analyzing power $A_y$ at 10 MeV, seemingly overcoming the long-standing $A_y$ puzzle. The price for this fit, however, is an increased discrepancy for the 
neutron-deuteron doublet scattering length $a_2$, as can be seen in Table~\ref{tab:u}.
The 3N potential U4 that improves $a_2$ compared to U1,
increases the discrepancy for $A_y$ at the same time, much like U2 and U3 do.
Thus again, the simultaneous account of 2N data, $ |E_{^{3} \rm H}|$,  $a_2$, and $A_y$
appears to be impossible with the presently available force models.
The models CD Bonn + $\Delta$ + U4 and  CD Bonn + $\Delta$ + U5, 
although the latter is successful for low-energy $A_y$,
contain  unusually large parameters  $A_{2\pi}$ and
$C_{2\pi}$, which make the 3N potentials U4 and U5
as questionable as U3 before. As shown in Fig.~\ref{fig:nd135u5}
for proton-deuteron elastic scattering at 135 MeV, especially U5
yields a very strong effect that destroys the reasonable agreement with data 
achieved by  the potentials CD Bonn + $\Delta$  or  CD Bonn + $\Delta$ + U1.

\begin{figure}[!]
\begin{center}
\includegraphics[scale=0.64]{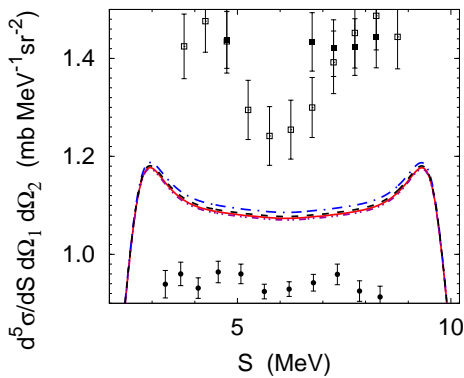}
\end{center}
\caption{\label{fig:nd13} (Color online) Differential cross section
  for neutron-deuteron breakup at 13~MeV neutron lab energy  in the space star
  configuration as function of the arclength $S$ along the kinematical
  curve.  Results  CD Bonn + $\Delta$
  (dashed curve),  CD Bonn + $\Delta$ + U1 (solid curve), CD Bonn +
  $\Delta$ + U3 (dotted-dashed curve), and CD Bonn +
  $\Delta$ + U5  (double-dotted-dashed curve) are compared with the
  experimental data from Refs.~\cite{strate:89,setze:05a} (open and
  full squares).  The proton-deuteron data from Ref.~\cite{rauprich:91} (full circles)
  are also shown.}
\end{figure}

Finally, Fig.~\ref{fig:nd13} studies another famous 3N puzzle, as bothersome 
as the $A_y$ problem, i.e., the fivefold differential cross section
$d^5\sigma/dS d\Omega_1 d\Omega_2$ of the nucleon-deuteron breakup at 13 MeV in the
space star configuration characterized by the polar and azimuthal
angles of two detected nucleons 
$(\theta_1, \theta_2, \varphi_2 - \varphi_1)
=(50.5^\circ,50.5^\circ,120.0^\circ)$.
This observable is known to be quite insensitive to changes in the Hamiltonian
\cite{witala:10a,ishikawa:09a}.
Indeed, even the effect of the very strong 3N forces U3 and U5 is insignificant and 
much smaller than the discrepancy between the predictions and experimental data
\cite{strate:89,setze:05a,rauprich:91}. Fig.~\ref{fig:nd13} presents theoretical 
results for neutron-deuteron scattering; the corresponding theoretical results 
for proton-deuteron breakup also strongly disagree with the data, despite the 
inclusion of Coulomb in the calculations; the proton-deuteron results overpredict the data.
Thus, none of the considered 3N force models studied by us is able to cure the
space star anomaly.}

{
\subsection{Overview on results obtained with the additional 3N potential U1} }

\begin{figure}[!]
\begin{center}
\includegraphics[scale=0.58]{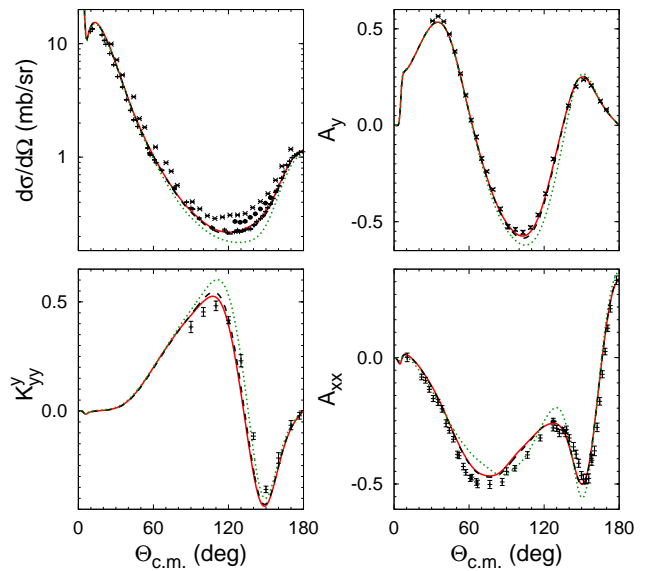}
\end{center}
\caption{\label{fig:pd135} (Color online) Differential cross section,
  proton analyzing power $A_y$, deuteron-to-proton spin-transfer coefficient
$K_{yy}^y$,  and deuteron analyzing power $A_{xx}$ 
for proton-deuteron elastic
  scattering at  135~MeV proton energy.  Results based on the
  potentials CD Bonn (dotted curve), CD Bonn + $\Delta$ (dashed curve)
  and  CD Bonn + $\Delta$ + U1 (solid curve) are shown.  
The  experimental data as in Fig.~\ref{fig:nd135u3}.}
\end{figure}

\begin{figure}[!]
\begin{center}
\includegraphics[scale=0.58]{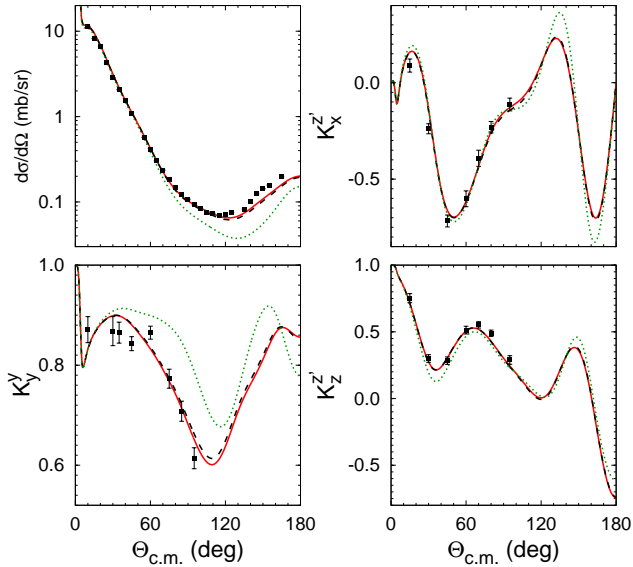}
\end{center}
\caption{\label{fig:pd250} (Color online) Differential cross section
   and nucleon-to-nucleon spin-transfer coefficients for
  proton-deuteron elastic scattering at  250~MeV proton energy. 
Note that this energy is slightly above the  $\pi$-production threshold
 but remains well below the $\Delta$-production threshold.
  Curves as in Fig.~\ref{fig:pd135}.  The experimental data are from
  Ref.~\cite{hatanaka:02a}.}
\end{figure}

\begin{figure}[!]
\begin{center}
\includegraphics[scale=0.62]{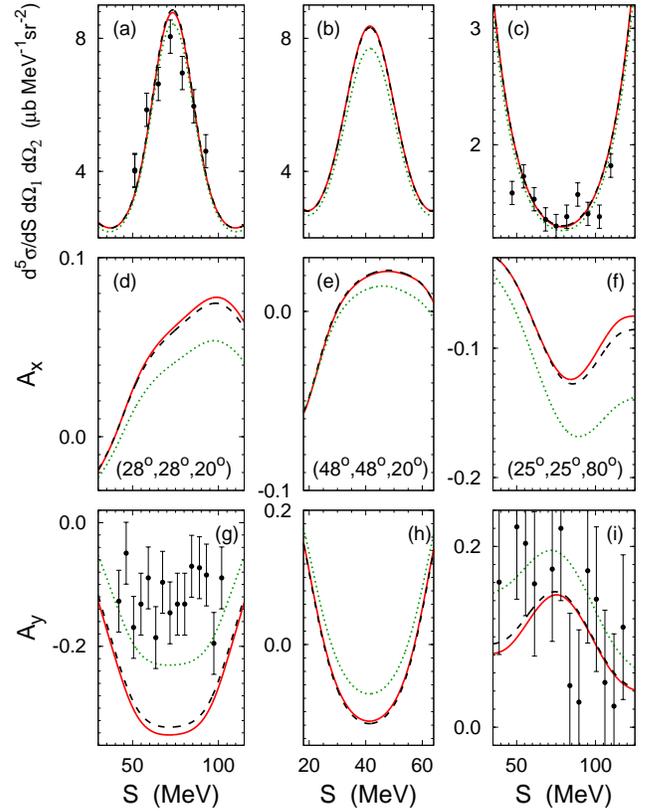}
\end{center}
\caption{\label{fig:pd135b} { (Color online) Differential cross section and
  proton analyzing powers $A_x$ and $A_y$
for proton-deuteron breakup  at  135~MeV proton energy.  
Results for three kinematical configurations 
$(28^\circ,28^\circ,20^\circ)$, $(48^\circ,48^\circ,20^\circ)$ and $(25^\circ,25^\circ,80^\circ)$ are
given from left and to right, respectively.
Curves as in Fig.~\ref{fig:pd135}.
The  experimental data are from  Ref.~\cite{eslami:10a}.}}
\end{figure}

We choose the irreducible 3N potential U1 as the most
reasonable one for a further comparative study of irreducible and
reducible contributions to the 3N force; the addition of U1 accounts
for the binding of $^{3} \rm H$ by construction, but it also yields
the  $^{3} \rm H$ - $^{3} \rm He$ 
mass difference perfectly, i.e., the binding of $^{3}\rm He$
 becomes 7.718 MeV, its experimental value also being 7.718
MeV. In this subsection for a wide range of 3N scattering
observables we compare the predictions of
the potentials CD Bonn without any 3N force, CD Bonn +
$\Delta$ with reducible contributions to the 3N force, and of CD
Bonn + $\Delta$ + U1 with reducible and irreducible contributions to
the 3N force. 

In  Fig.~\ref{fig:pd135}
we show the differential cross section and various spin observables
for elastic proton-deuteron scattering at 135 MeV, in 
Fig.~\ref{fig:pd250} the differential cross section and
nucleon-to-nucleon spin transfer coefficients 
for elastic proton-deuteron scattering at 250 MeV,
and in Fig.~\ref{fig:pd135b} the fivefold differential cross section
and nucleon analyzing powers for  proton-deuteron breakup at 135 MeV.
In contrast to Subsection III.A, we carry out the calculations for
proton-deuteron scattering fully and include the proton-proton Coulomb force
using the method of screening and renormalization
\cite{taylor:74a,alt:80a,deltuva:05a,deltuva:05d}, though we do not show the explicit Coulomb effects
separately in the plots. 
In elastic scattering at these energies the Coulomb effect is irrelevant,
except for small angles \cite{deltuva:05a};
it is seen in Figs.~\ref{fig:pd135} and Fig.~\ref{fig:pd250}
as wiggles in all observables below $\Theta_\cm < 10^\circ$.
In breakup the Coulomb effect is most visible in configurations with
low relative proton-proton energy \cite{deltuva:05d}, mostly affecting the differential
cross section. For example, in the configuration
$(28^\circ,28^\circ,20^\circ)$ of Fig.~\ref{fig:pd135b} the proton-proton Coulomb
force reduces the cross section at the peak by about 25\%.

The reducible contributions to the 3N force, i.e., the explicit 
$\Delta$-isobar excitation,
create the largest effects in the studied observables. The additional
effects of the irreducible 3N force U1 are minor even at higher energies, and they
do not destroy any achieved account of the experimental data. In  
proton-deuteron elastic scattering,
the reducible and irreducible 3N-force contributions move the predictions quite often into the same direction
such that the inclusion of irreducible 3N force U1 even slightly improves the 
description of the experimental data. The most prominent examples are
the differential cross section around the minimum and at backward angles, the nucleon
analyzing power $A_y$ at 135 MeV around the minimum and the nucleon-to-nucleon
spin transfer coefficient $K_y^y$ at 250 MeV, as well as 
 the deuteron vector analyzing power $A_y(d)$  at 135 MeV shown 
in Fig.~\ref{fig:nd135u3}.
In the breakup reaction
the reducible and irreducible 3N forces may have effects of the same or of opposite 
signs as illustrated in  Fig.~\ref{fig:pd135b}, depending on the kinematical configuration and depending on 
the specific observable. In the considered example the Coulomb 
repulsion between the protons is important for the achieved gross agreement with the 
cross section data, but 
neither 3N force contribution is beneficial in accounting for the experimental $A_y$
data  of Ref.~\cite{eslami:10a}, especially in the $(28^\circ,28^\circ,20^\circ)$ configuration.

\section {Conclusions}

In the present work we describe the properties of the three-nucleon system, i.e.,
the bound state and the nucleon-deuteron elastic scattering and breakup below
the  $\pi$-production threshold, in an extended Hilbert space with a coupled-channel 
Hamiltonian containing irreducible two- and three-baryon
potentials. The extension of the Hilbert space is due to a sector in
which one nucleon is turned into a $\Delta$ isobar. The Hamiltonian couples both
 sectors of the Hilbert space. The $\Delta$
isobar makes important, reducible, easily calculable contributions to
the 2N and 3N interactions; the most prominent contribution is the
Fujita-Miyazawa mechanism to the 3N interaction, even enriched by the $\rho$ meson 
contributions. The irreducible
three-baryon potential is constrained to a purely nucleonic one. The
simplification of the additional Hilbert sector with a single $\Delta$
isobar only and the constraint of the irreducible three-baryon
potential to a nucleonic one only are well motivated. 

We prove that calculations in such an extended Hilbert space
with such an extended Hamiltonian can be computationally done with
high precision. The computational technology can, in future, be
extended to other and conceptually better founded forces, e.g., to
those possibly provided by $\chi$EFT  with an explicit $\Delta$ 
isobar ~\cite{krebs:07,krebs:08}, kept 
in a quantum-mechanical  Hamiltonian still to be derived. Thus, the technical
objective of the paper is reached.  

We  explored the effect  of irreducible 3N potentials of the
Urbana IX type on top of the  coupled-channel two-baryon potential CD
Bonn + $\Delta$.  We found that the simultaneous precise account of
the  $^{3} \rm H$ binding energy $ |E_{^{3} \rm H}|$ and of the neutron-deuteron doublet 
scattering length $a_2$ is extremely difficult to achieve. This
finding may appear to be due to our special  theoretical
framework. But this is a finding encountered  also in Ref.~\cite{epelbaum:14}
using the framework of  entirely different nuclear forces
of  $\chi$EFT  and in Ref.~\cite{kievsky:10a}  employing the Argonne V18 potential 
combined with variations of the Urbana 
IX and  Tucson-Melbourne 3N forces. 
Thus, this finding is surprisingly more general than
expected and may  constitute another theoretical puzzle in the
three-nucleon system.  
Furthermore, the simultaneous account of 2N data, of the $^{3}\rm H$ binding 
$ |E_{^{3} \rm H}|$, of the neutron-deuteron doublet  scattering length $a_2$ and 
of the low-energy nucleon analyzing power 
$A_y$ appears to be even more difficult to achieve, and that problem also 
calls for new  terms in the Hamiltonian, especially of the 3N nature.

We employed  the irreducible 3N potential U1 of  Table
\ref{tab:u} as dynamic basis for the description of three-nucleon
binding and scattering,  together with the coupled-channel two-baryon
potential CD Bonn + $\Delta$.  The 3N potential U1 provides the
missing contribution to the three-nucleon  binding and it is weak.
This is wanted, but we have to give up our original goal of also
accounting for the neutron-deuteron doublet  scattering length $a_2$
precisely.   
Nevertheless, the results for three-nucleon scattering at higher energies are
satisfactory, since the broad range of the successful account of
experimental data without an irreducible  3N potential is preserved
and often even slightly improved. 
However, the well-known puzzles of low-energy three-nucleon scattering,  i.e., the  nucleon analyzing
power $A_y$ of elastic nucleon-deuteron scattering below 20 MeV nucleon 
lab energy  and the nucleon-deuteron
breakup in the space star kinematics stay unresolved.
That is a disappointing aspect of the present results, but not 
unexpected given the previous studies \cite{kievsky:10a,epelbaum:14,witala:10a,ishikawa:09a}.

\begin{acknowledgments}
A.D. thanks the Leibniz University Hannover for the hospitality during
the completion of this work.
\end{acknowledgments}


\begin{thebibliography}{10}

\bibitem{machleidt:87a}
R. Machleidt, K. Holinde, and C. Elster, Phys.~Rep. {\bf 149},  1  (1987).

\bibitem{epelbaum:06} E. Epelbaum, Prog.~Part.~Nucl.~Phys. {\bf 57},
  654 (2006).

\bibitem{epelbaum:09} E. Epelbaum, H.-W. Hammer and U.-G. Mei{\ss}ner,
  Rev. Mod. Phys. {\bf 81}, 1773 (2009).

\bibitem{machleidt:11a}
R. Machleidt and D.~R. Entem, Phys. Rep. {\bf 503},  1  (2011).

\bibitem{krebs:07} H. Krebs, E. Epelbaum and U.-G. Mei{\ss}ner,
  Eur.~Phys.~J. A{\bf 32}, 127 (2007).

\bibitem{krebs:08} E. Epelbaum, H. Krebs and U.-G. Mei{\ss}ner,
  Nucl.~Phys.~A{\bf 806}, 65 (2008).

\bibitem{deltuva:03c}
A. Deltuva, R. Machleidt, and P.~U. Sauer, Phys.~Rev.~C {\bf 68},  024005
  (2003).

\bibitem{fujita:57a}
J. Fujita and H. Miyazawa, Prog.~Theor.~Phys. {\bf 17},  360  (1957).

\bibitem{krebs:13} H. Krebs, A. Gasparyan and E. Epelbaum,
  Phys.~Rev.~C{\bf 87}, 054007 (2013)

\bibitem{pieper:01a}
S.~C. Pieper, V.~R. Pandharipande, R.~B. Wiringa, and J. Carlson, Phys.~Rev.~C
  {\bf 64},  014001  (2001).

\bibitem{nemoto:98a}
S. Nemoto, K. Chmielewski, J. Haidenbauer, S. Oryu, P.~U. Sauer, and N.~W.
  Schellingerhout, Few-Body Systems {\bf 24},  213  (1998).

\bibitem{deltuva:03a}
A. Deltuva, K. Chmielewski, and P.~U. Sauer, Phys.~Rev.~C {\bf 67},  034001
  (2003).

\bibitem{deltuva:09e}
A. Deltuva, Phys.~Rev.~C {\bf 80},  064002  (2009).

\bibitem{stadler:92a}
A. Stadler and P.~U. Sauer, Phys. Rev. C {\bf 46},  64  (1992).

\bibitem{alt:67a}
E.~O. Alt, P. Grassberger, and W. Sandhas, Nucl.~Phys. {\bf B2},  167  (1967).

\bibitem{urbana9}
B.~S. Pudliner, V.~R. Pandharipande, J. Carlson, S.~C. Pieper, and R.~B.
  Wiringa, Phys.~Rev. C {\bf 56},  1720  (1997).

\bibitem{lazauskas:phd}
R. Lazauskas, Ph.D. thesis, Universit\'{e} Joseph Fourier, Grenoble, 2003, 
  http://tel.ccsd.cnrs.fr/documents/archives0/00/00/41/78/.

\bibitem{taylor:74a}
J.~R. Taylor, Nuovo Cimento B {\bf 23},  313  (1974); M.~D. Semon and J.~R.
  Taylor, Nuovo Cimento A {\bf 26}, 48 (1975).

\bibitem{alt:80a}
E.~O. Alt and W. Sandhas, Phys.~Rev.~C {\bf 21},  1733  (1980).

\bibitem{deltuva:05a}
A. Deltuva, A.~C. Fonseca, and P.~U. Sauer, Phys.~Rev.~C {\bf 71},  054005
  (2005).

\bibitem{deltuva:05d}
A. Deltuva, A.~C. Fonseca, and P.~U. Sauer, Phys.~Rev.~C {\bf 72},  054004
  (2005).

\bibitem{wiringa:95a}
R.~B. Wiringa, V.~G.~J. Stoks, and R. Schiavilla, Phys.~Rev.~C {\bf 51},  38
  (1995).

\bibitem{coon:79}
S.~A. Coon, M.~D. Scadron, P.~C. McNamee, B.~R. Barrett, D.~W.~E. Blatt, and B.~H.~J. McKellar,
Nucl.~Phys.~A {\bf 317}, 242 (1979). 

\bibitem{manzke:78}
W. Manzke and M. Gari, Nucl. Phys. A {\bf 312},  457   (1978).

\bibitem{dilg:71}
W. Dilg, L. Koester, and W. Nistler, Phys.~Lett. {\bf 36B},  208  (1971).

\bibitem{kievsky:10a} 
A. Kievsky, M. Viviani, L. Girlanda, and L.E. Marcucci, Phys.~Rev.~C {\bf 81},  044003
  (2010).

  
\bibitem{epelbaum:14} J. Golak et al, 
Eur. Phys. J. A  {\bf 50}, 177 (2014).

\bibitem{witala:03a}
H. Wita\l{}a, A. Nogga, H. Kamada, W. Gl\"ockle, J. Golak, and R.
  Skibi\ifmmode~\acute{n}\else \'{n}\fi{}ski, Phys. Rev. C {\bf 68},  034002
  (2003).

\bibitem{howell:87}
C.~R. Howell {\it et~al.}, Few-Body Systems {\bf 2},  19  (1987).

\bibitem{ermisch:05a}
K. Ermisch {\it et~al.}, Phys. Rev. C {\bf 71},  064004  (2005).

\bibitem{ramazani:08a}
A. Ramazani-Moghaddam-Arani {\it et~al.}, Phys. Rev. C {\bf 78},  014006
  (2008).

\bibitem{sekiguchi:02a}
K. Sekiguchi {\it et~al.}, Phys.~Rev.~C {\bf 65},  034003  (2002).

\bibitem{witala:10a}
H. Wita{\l}a and W. Gl\"{o}ckle, J. Phys. G {\bf 37},  064003  (2010).

\bibitem{ishikawa:09a}
S. Ishikawa, Phys. Rev. C {\bf 80},  054002  (2009).

\bibitem{strate:89}
J. Strate {\it et~al.}, Nucl.~Phys. {\bf A501},  51  (1989).

\bibitem{setze:05a}
H.~R. Setze~{\it et al}, Phys.~Rev.~C {\bf 71},  034006  (2005).

\bibitem{rauprich:91}
G. Rauprich, S. Lemaitre, P. Niessen, K.~R. Nyga, R. Reckenfelderb\"aumer, L.
  Sydow, H. Paetz~gen. Schieck, H. Wita{\l}a, and W. Gl\"ockle, Nucl.~Phys.
  {\bf A535},  313  (1991).

\bibitem{hatanaka:02a}
K. Hatanaka~{\it et al.}, Phys.~Rev.~C {\bf 66},  044002  (2002).

\bibitem{eslami:10a}
M. Eslami-Kalantari~{\it et al}, EPJ Web of Conferences {\bf 3},  05010
  (2010).

\end{thebibliography}


\end{document}